\documentclass[aps,prd,twocolumn,floatfix,nofootinbib,showpacs,superscriptaddress]{revtex4-1}

\usepackage{latexsym}
\usepackage{amsmath}
\usepackage{amssymb}
\usepackage{graphicx}
\usepackage{longtable}

\usepackage{bbm}
\usepackage{epsfig}
\usepackage[usenames]{color}

\begin{document}

\newcommand{\dslash}[1]{#1 \hspace{-1.2ex}\slash}

\title{Existence and stability of multiple solutions to the gap equation}

\author{Kun-lun Wang}
\affiliation{Department of Physics, Center for High Energy Physics and State Key Laboratory of Nuclear Physics and Technology, Peking University, Beijing 100871, China}

\author{Si-xue Qin}
\affiliation{Department of Physics, Center for High Energy Physics and State Key Laboratory of Nuclear Physics and Technology, Peking University, Beijing 100871, China}

\author{Yu-xin Liu}
\email[Corresponding author: ]{yxliu@pku.edu.cn}
\affiliation{Department of Physics, Center for High Energy Physics and State Key Laboratory of Nuclear Physics and Technology, Peking University, Beijing 100871, China}

\author{Lei Chang}
\affiliation{Institut f\"ur Kernphysik, Forschungszentrum J\"ulich, D-52425 J\"ulich, Germany}

\author{Craig D.~Roberts}
\email[Corresponding author: ]{cdroberts@anl.gov}
\affiliation{Institut f\"ur Kernphysik, Forschungszentrum J\"ulich, D-52425 J\"ulich, Germany}
\affiliation{Physics Division, Argonne National Laboratory, Argonne, Illinois 60439, USA}
\affiliation{Department of Physics, Illinois Institute of Technology, Chicago, Illinois 60616-3793, USA}

\author{Sebastian M.~Schmidt}
\affiliation{Institute for Advanced Simulation, Forschungszentrum J\"ulich and JARA, D-52425 J\"ulich, Germany}

\date{28 August 2012}

\begin{abstract}
We argue by way of examples that, as a nonlinear integral equation, the gap equation can and does possess many physically distinct solutions for the dressed-quark propagator.  The examples are drawn from a class that is successful in describing a broad range of hadron physics observables.  We apply the homotopy continuation method to each of our four exemplars and thereby find all solutions that exist within the interesting domains of light current-quark masses and interaction strengths; and simultaneously provide an explanation of the nature and number of the solutions, many of which may be associated with dynamical chiral symmetry breaking.  Introducing a stability criterion based on the scalar and pseudoscalar susceptibilities we demonstrate, however, that for any nonzero current-quark mass only the regular Nambu solution of the gap equation is stable against perturbations.  This guarantees that the existence of multiple solutions to the gap equation cannot complicate the description of phenomena in hadron physics.
\end{abstract}

\maketitle

\section{Introduction}
%
Dynamical chiral symmetry breaking (DCSB) is a particularly striking feature of the Standard Model, playing an important role in formation of the visible matter in the Universe \cite{national2012Nuclear}.
%
%
It is apparent in the existence of a strongly momentum-dependent chiral-limit dressed-quark mass function, $M(p^2)$, which is obtained in solutions of models for QCD's gap equation that provide a realistic description of hadron properties \cite{Chang:2011vu,Bashir:2012fs,Roberts:2012sv}, and in a sharp increase in $M(p^2)$ at $p^2 \lesssim 2\,$GeV$^2$ when the current-quark mass is nonzero but light.  The latter is seen in both Dyson-Schwinger equation (DSE) studies \cite{Bhagwat:2003vw,Bhagwat:2004kj,Roberts:2007ji} and numerical simulations of lattice-regularised QCD \cite{Bowman:2002bm,Bowman:2004jm,Kamleh:2007ud}.  This behaviour of the mass function must be part of any treatment of continuum strong QCD that aims to be considered realistic.
%

The gap equation is a nonlinear integral equation.  The nonlinearity gives it the power to express nonperturbative phenomena; and also leads to the curiosity that the solution is not unique.  Mathematically, this should have been anticipated.  However, perhaps surprisingly, it has only been established relatively recently that on a bounded, measurable domain of non-negative current-quark mass, realistic models of the QCD's gap equation simultaneously admit more than one nonequivalent DCSB solution and also distinct solutions that may unambiguously be connected with the realization of chiral symmetry in the Wigner mode \cite{Zong:2004nm,Chang:2006bm,Williams:2006vva}.  This feature can potentially create problems; e.g., if the additional solutions are physically realisable and therefore have a measurable impact on observables.  Thus, amongst the questions that must be answered is that of which solution or solutions of the gap equation should be employed in defining the kernel of the Bethe-Salpeter two-body problem.  Naturally, in addressing such questions, the first thing to ensure is that one has found all solutions to the gap equation.

Herein we describe solutions of the gap equation obtained using two different interaction models \cite{Maris:1999nt,Qin:2011dd} and two vertex \emph{Ans\"atze}.  We focus on the interesting domain of light current-quark masses and a large domain of interaction strengths.  Notably, we employ a numerical method \cite{Homotopy}, novel in the study of DSEs, which delivers all solutions of the gap equation.  This enables us to chart the complete solution set domains for each of the four kernels we consider.  We also discuss the important issue of stability for each of the various solutions and thereby address the question raised above.

Section~\ref{sec:gapeqn} provides a brief review of the gap and Bethe-Salpeter equations.  It also explains that all model-independent content of the so-called ``Mexican hat'' potential is encoded in the behaviour of the scalar and pseudoscalar susceptibilities, which therefore provide a general tool for judging the stability of gap equation solutions, insofar as any given solution might represent a valid starting point in the computation of hadron properties.  Section~\ref{sec:models} introduces the kernel \emph{Ans\"atze} that we employ.  They are not new but were chosen because they are capable of describing and unifying a wide range of meson and baryon properties.  Section~\ref{sec:results} is an extensive presentation and discussion of the solutions to the gap equation, which explains: their classification; their evolution in response to changes in two control parameters (current-quark mass and interaction strength); and the process of identifying stable solutions.  Section~\ref{sec:epilogue} is a recapitulation with some additional comments.  We explain our numerical method for solving the gap equation in App.\,\ref{app:A1}.

\section{Gap equation}
\label{sec:gapeqn}
Since DCSB is a phenomenon emerging from the strong physics of dressed-quarks, it is often studied via QCD's gap equation:\footnote{We use a Euclidean metric:  $\{\gamma_\mu,\gamma_\nu\} = 2\delta_{\mu\nu}$; $\gamma_\mu^\dagger = \gamma_\mu$; $\gamma_5= \gamma_4\gamma_1\gamma_2\gamma_3$, tr$[\gamma_5\gamma_\mu\gamma_\nu\gamma_\rho\gamma_\sigma]=-4 \epsilon_{\mu\nu\rho\sigma}$; $\sigma_{\mu\nu}=(i/2)[\gamma_\mu,\gamma_\nu]$; $a \cdot b = \sum_{i=1}^4 a_i b_i$; and $P_\mu$ timelike $\Rightarrow$ $P^2<0$.}
\begin{eqnarray}
\lefteqn{
\nonumber S^{-1}(p) = Z_2 \,(i\gamma\cdot p + m^{\rm bm})}\\
&& + Z_1 \int^\Lambda_{dq}\!\! g^2 D_{\mu\nu}(p-q)\frac{\lambda^a}{2}\gamma_\mu S(q) \frac{\lambda^a}{2}\Gamma_\nu(q,p) ,
\label{gendseN}
\end{eqnarray}
where: $D_{\mu\nu}$ is the gluon propagator; $\Gamma_\nu$, the quark-gluon vertex; $\int^\Lambda_{dq}$, a symbol representing a Poincar\'e invariant regularisation of the four-dimensional integral, with $\Lambda$ the regularisation mass-scale; $m^{\rm bm}(\Lambda)$, the current-quark bare mass; and $Z_{1,2}(\zeta^2,\Lambda^2)$, respectively, the vertex and quark wave-function renormalisation constants, with $\zeta$ the renormalisation point.  
We employ the renormalisation procedures of Ref.\,\cite{Maris:1997tm} and the same renormalisation point, $\zeta=19\,$GeV.

The gap equation's solution is the dressed-quark propagator, which is commonly written in one of three equivalent forms:
\begin{subequations}
\begin{eqnarray}
S(p)
& = & -i \gamma\cdot p \, \sigma_V(p^2,\zeta^2)+\sigma_S(p^2,\zeta^2)\,, \label{SgeneralSigma}\\
&=& 1/[i \gamma\cdot p \, A(p^2,\zeta^2) + B(p^2,\zeta^2)]\,,
\label{SgeneralNAB}\\
&=&Z(p^2,\zeta^2)/[i\gamma\cdot p + M(p^2)]\,.
\label{SgeneralN}
\end{eqnarray}
\end{subequations}
The mass function, $M(p^2)$, is independent of the renormalisation point; and the renormalised current-quark mass,
\begin{equation}
\label{mzeta}
m^\zeta = Z_m(\zeta,\Lambda) \, m^{\rm bm}(\Lambda) = Z_4^{-1} Z_2\, m^{\rm bm},
\end{equation}
where $Z_4$ is the renormalisation constant associated with the Lagrangian's mass-term.  The renormalisation-group invariant current-quark mass may be inferred via
\begin{equation}
\hat m = \lim_{p^2\to\infty} \left[\frac{1}{2}\ln \frac{p^2}{\Lambda^2_{\rm QCD}}\right]^{\gamma_m} M(p^2)\,,
\label{RGIcqmass}
\end{equation}
where $\gamma_m = 12/(33-2 N_f)$ with $N_f$ the number of active quark flavours.  The chiral limit is
\begin{equation}
\hat m = 0\,.
\end{equation}

Chiral-limit QCD possesses a $SU_L(N_f)\otimes SU_R(N_f)$ symmetry and thus separates into two non-communicating theories: one for left-handed quarks and another for right-handed quarks.  This can be seen to entail that the regular parts of the scalar and pseudoscalar vacuum susceptibilities must be identical \cite{Chang:2009at}.  In fact, this is the content of the so-called ``Mexican hat'' potential, which is commonly used in building models for QCD.

The symmetry requires that the gap equation is invariant under a change in the sign of $B(p^2)$ in Eq.\,\eqref{SgeneralNAB}; i.e., if $B_0$ is a solution, then so is $(-B_0)$.  In the context of simple gap equation truncations, this has long been known \cite{Cahill:1985mh,Roberts:1985ju}.  On the other hand, as we now describe, it limits what may be called realistic \emph{Ans\"atze} for the dressed-quark-gluon vertex.  The dressed vertex separates into a sum of two pieces: one in which every term is even in the number of Dirac matrices, $\Gamma_\mu^{\rm D-even}$; and another in which every term is odd, $\Gamma_\mu^{\rm D-odd}$.  The vertex satisfies its own DSE, the kernel of which features the dressed-quark propagator.  The result described above entails that $\Gamma_\mu^{\rm D-even}$ obtained as a solution of this DSE must change sign under $B \to -B$ but is otherwise unchanged, whereas $\Gamma_\mu^{\rm D-odd}$ is invariant under $B \to -B$.  This is a significant result because the scalar functions that accompany the even and odd tensor structures in the vertex cannot in general be expected to share the functional form of either $A(p^2)$ or $B(p^2)$, or simple combinations thereof.  Notwithstanding this, most vertex models are constructed with a simple functional dependence on $A(p^2)$ and $B(p^2)$, such that they do exhibit the correct properties under $B \to -B$ \cite{Ball:1980ay,Curtis:1990zs,Kizilersu:2009kg,Chang:2010hb,Chang:2011ei,Bashir:2011dp}.

In the chiral limit, therefore, at least two possibilities are realisable.  Namely, if the support of the integrand in the equation for $B(p^2)$ is too small, then the gap equation possesses solely the $B=B_W\equiv 0$ solution, whereas, if the support exceeds some critical value, it has three solutions; viz., $B_W$, $B=B^+_{N}=B_0$, $B=B^-_{N}=-B_0$.
In studies that convert the gap equation into a second-order nonlinear differential equation for $B(p^2)$ \cite{Atkinson:1988mw,Roberts:1989mj}, a step which is quantitatively accurate for $p^2\gtrsim 2\,$GeV$^2$, it is natural to characterise the latter two solutions as regular, whereas the first solution can be connected with an irregular solution.
It is more common, however, to denominate the $B\equiv 0$ solution as the Wigner-Weyl mode and the latter two as Nambu-Goldstone type solutions, since they signal the dynamical breaking of chiral symmetry.

It is notable that in the chiral limit the Nambu solutions are energetically favoured in concrete computations that produce both the Wigner- and Nambu-type solutions \cite{Qin:2010nq}.  In the context of Ref.\,\cite{Nambu:1961tp}, the continuum of Nambu solutions would be described as equivalent degenerate vacua.  If one introduces a small current-quark mass, then solutions smoothly connected to $B_N^+$, $B_N^-$ and $B_W$ persist \cite{Zong:2004nm,Chang:2006bm,Williams:2006vva}.  Plainly, therefore, mere existence as a solution of the gap equation does not guarantee that solution's stability.

In investigating stability of solutions to the gap equation it has proved useful to employ the chiral susceptibility, which is defined as usual via the scalar vacuum polarisation.  A solution is energetically unstable in response to fluctuations of some source if the associated chiral susceptibility is negative \cite{Yuan:2006yf,Zhao:2008zzi,Chang:2008ec,Qin:2010nq}.  The information is incomplete, however, since if the susceptibility is positive semi-definite, then the solution may be stable, meta-stable, or a saddle-point.  The last possibility is real here because the scalar and pseudoscalar vacuum polarisations are distinguishable when chiral symmetry is dynamically broken \cite{Chang:2009at}, and the pseudoscalar susceptibility can be negative whilst the scalar susceptibility is positive semi-definite.

\begin{figure}[t]
\includegraphics[width=1.0\linewidth]{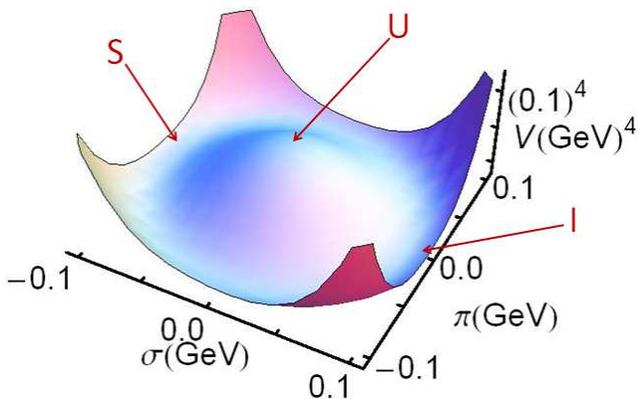}
\caption{Classical potential often imagined in connection with dynamical chiral symmetry breaking.  The point ``I'' is the global minimum, characterised by the conditions in Eqs.\,\protect\eqref{d2Usp}; ``S'' is a saddle-point, $m_s^2>0$ but $m_p^2<0$; and $U$ is an unstable local maximum $m_s^2<0$ and $m_p^2<0$.
\label{fig:MexHat}}
\end{figure}

These polarisations
are obtained via second-order functional derivatives of the theory's generating functional for connected one-particle-irreducible Schwinger functions; viz., $\delta^2 \Gamma_{\rm 1PI}/\delta J(x) \delta J(y)$, where $J=S,P$ are respectively scalar and pseudoscalar sources. 
Their importance and relevance herein are evident once one appreciates that a simultaneous consideration of the scalar and pseudoscalar vacuum polarisations expresses, amongst other things, all model-independent physical content of the so-called ``Mexican hat'' potential \cite{Chang:2008ec}.  Using this connection here for illustrative simplicity, suppose that potential is represented as $U[s,p]$.  In this case an extremum $\vec{v}=\{\bar s,\bar p\}$, is stable if, and only if,
\begin{subequations}
\label{d2Usp}
\begin{eqnarray}
&& m_s^2 := \frac{1}{2}\left. \frac{\partial^2}{\partial s^2}U[s,p] \right|_{{\bar s,\bar p}}>0\,,\\
&& m_p^2 := \frac{1}{2}\left. \frac{\partial^2}{\partial p^2}U[s,p] \right|_{{\bar s,\bar p}}>0\,.
\end{eqnarray}
\end{subequations}
If just one of $m_s^2$, $m_p^2$ is negative, then $\vec{v}$ is a saddle-point; whereas if both are negative, then $\vec{v}$ is a local maximum.  These points are depicted in Fig.\,\ref{fig:MexHat}.
%

Translating back to the general case, Eqs.\,\eqref{d2Usp} correspond to the statement that a solution configuration is truly stable if, and only if, both susceptibilities are positive at zero total momentum.  One can extend this by noting that each susceptibility is the inverse of a fully-dressed propagator for composite correlations in the relevant channel.  It follows that Eqs.\,\eqref{d2Usp} translate into the statement that stability is guaranteed if, and only if, the lowest mass excitation in each channel has positive mass-squared.  This is the criterion we use subsequently.  It can be implemented simply by solving the inhomogeneous scalar and pseudoscalar Bethe-Salpeter equations \cite{Chang:2009zb}:
\begin{eqnarray}
\nonumber
\lefteqn{\Gamma_J(q;K) = Z_4 M_J -\int_{d\ell}^\Lambda g^2 D_{\mu\nu}(q-\ell) } \\
& &
\nonumber
\times \frac{\lambda^a}{2}\gamma_\mu S(\ell_+) \Gamma_J(\ell;K) S(\ell_-) \frac{\lambda^a}{2} \Gamma_\nu(\ell_-,q_-)\\
&+&\int_{dq}^\Lambda g^2 D_{\mu\nu}(q-\ell )\frac{\lambda^a}{2} \gamma_\mu S(\ell_+)
\frac{\lambda^a}{2}\Lambda_{J\nu}(q,\ell;K) \label{genbse}
\end{eqnarray}
wherein the dressed-quark propagator is that which characterises the gap equation solution whose stability is in question and, furthermore:
$M_{\{S,P\}}=\{\mathbf{I}, \gamma_5\}$;
$\ell_\pm = \ell\pm K/2$, without loss of generality in our Poincar\'e covariant approach, where $K$ is the total momentum entering the vertex; $\Lambda_{J\nu}$ is a Bethe-Salpeter kernel, which is fully determined by the kernel of the gap equation; and the Bethe-Salpeter amplitudes have the general form
\begin{subequations}
\begin{eqnarray}
\Gamma_S(q;K) & = & [E_S(q;K)+ i \gamma\cdot K \,F_S(q;K)\\\nonumber
&+&  i\, \gamma\cdot q \, G_S(q;K) + \sigma_{\mu\nu}q_\mu K_\nu H_S(k;P)] \, ,\\
\Gamma_P(q;K) & = &\gamma_5 [ i E_P(q;K)+ \gamma\cdot K \, F_P(q;K)\\\nonumber
&+& \gamma\cdot q \, G_P(q;K)
+ \sigma_{\mu\nu} q_\mu K_\nu H_P(q;K)] \, ,
\label{GammaP}
\end{eqnarray}
\end{subequations}
with $E_J$, $F_{J}$, $G_{J}$, $H_{J}$ being scalar functions.  We locate the lowest-mass excitation using the method of Ref.\,\cite{Bhagwat:2007rj}, which simplifies computations by permitting one to employ solely spacelike momenta.

\section{Model Kernels}
\label{sec:models}
The gap equation's kernel is specified by the form used to express the contraction $Z_1 g^2 D_{\mu\nu}(k-q)\Gamma_\nu(q,p)$ in Eq.\,\eqref{gendseN}.  Herein we compare four kernels, which may all be introduced by first writing ($k=p-q$)
\begin{eqnarray}
\nonumber
\lefteqn{
Z_1 \, g^2 D_{\mu\nu}(k) \Gamma_\nu(q,p) = k^2 {\cal G}(k^2)
D^{\rm free}_{\mu\nu}(k) \Gamma^A_\nu(q,p)}\\
&=&   \left[ k^2 {\cal G}_{\rm IR}(k^2) + 4\pi \tilde\alpha_{\rm pQCD}(k^2) \right]
D^{\rm free}_{\mu\nu}(k) \Gamma^A_\nu(q,p) ,
\label{rainbowdse}
\end{eqnarray}
wherein: $D^{\rm free}_{\mu\nu}(k)$ is the free-gauge-boson propagator;\footnote{
We use Landau gauge, a choice made for many reasons \protect\cite{Bashir:2009fv,Bashir:2011vg,Bashir:2011dp}, for example, it is: a fixed point of the renormalisation group; that gauge for which sensitivity to model-dependent differences between \emph{Ans\"atze} for the fermion--gauge-boson vertex are least noticeable; and a covariant gauge, which is readily implemented in simulations of lattice regularised QCD (see, e.g., Refs.\,\protect\cite{Bowman:2002bm,Bowman:2005vx,Bowman:2004jm,%
Cucchieri:2007md,Bogolubsky:2009dc,Boucaud:2011ug,Cucchieri:2011um,Cucchieri:2011aa}, and citations therein and thereto).}
$\tilde\alpha_{\rm pQCD}(k^2)$ is a bounded, monotonically-decreasing regular continuation of the perturbative-QCD running coupling to all values of spacelike-$k^2$; ${\cal G}_{\rm IR}(k^2)$ is an \emph{Ansatz} for the interaction at infrared momenta: ${\cal G}_{\rm IR}(k^2)\ll \tilde\alpha_{\rm pQCD}(k^2)$ $\forall k^2\gtrsim 2\,$GeV$^2$; and $\Gamma^A_\nu(q,p)$ is an \emph{Ansatz} for that part of the dressed-quark-gluon vertex which cannot be absorbed into ${\cal G}(k^2)$.

In all instances we use \cite{Maris:1997tm}
\begin{equation}
4\pi \tilde\alpha_{\rm pQCD}(k^2)
= \frac{8 \pi^2 \gamma_m\,  {\cal F}(s)}{\ln [ \tau + (1+s/\Lambda_{\rm QCD}^2)^2]},
\label{alphatilde}
\end{equation}
where: $\gamma_m = 12/(33-2 N_f)$, $N_f=4$, $\Lambda_{\rm QCD}=0.234\,$GeV; $\tau={\rm e}^2-1$; and ${\cal F}(s) = \{1 - \exp(-s/[4 m_t^2])\}/s$, $m_t=0.5\,$GeV.  For the infrared, we compare two forms; viz., \cite{Maris:1999nt,Qin:2011dd}
\begin{subequations}
\label{CalG}
\begin{eqnarray}
\label{CalGMT}
{\cal G}_{\rm IR}^{\rm MT}(s) &=& \frac{4 \pi^2}{\omega^6} D s \, {\rm e}^{-s/\omega^2} \,, \\
{\cal G}_{\rm IR}^{\rm QC}(s) &=& \frac{8 \pi^2}{\omega^4} D \, {\rm e}^{-s/\omega^2}\,.
\label{CalGQC}
\end{eqnarray}
\end{subequations}
These are actually one-parameter models because in both cases there is a domain of $\omega$ throughout which, in rainbow-ladder truncation -- see below, computed properties of ground state vector and flavour-nonsinglet pseudoscalar mesons \cite{Maris:1999nt,Maris:2002mt,Maris:2003vk}, and nucleon and $\Delta$ properties \cite{Eichmann:2008ef,Eichmann:2011ej} are almost unchanged along the trajectory $D\omega = \,$constant.
That domain is $\omega\in[0.3,0.5]\,$GeV for the interaction in Eq.\,\eqref{CalGMT}, whereupon $D\omega = (0.72\,{\rm GeV})^3$ provides a good description of the observables identified \cite{Maris:2002mt}; whilst for Eq.\,\eqref{CalGQC} the domain is $\omega\in[0.4,0.6]\,$GeV, with $D\omega = (0.8\,{\rm GeV})^3$ providing the best achievable phenomenological results \cite{Qin:2011dd}.  We will use the midpoint of each domain for computations throughout; i.e., $\omega=0.4\,$GeV in Eq.\,\eqref{CalGMT} and $\omega=0.5\,$GeV in Eq.\,\eqref{CalGQC}.

We employ two simple models for the vertex:
\begin{subequations}
\label{CalV}
\begin{eqnarray}
\label{CalVRL} \Gamma_\nu^{\rm RL}(q,p) & = & \gamma_\nu\,,\\
\label{CalV1BC} \Gamma_\nu^{\rm 1BC}(q,p) & = & \gamma_\nu \frac{A(q^2)+A(p^2)}{2}\,.
\end{eqnarray}
\end{subequations}
The first implements a rainbow-ladder truncation of the gap and Bethe-Salpeter equations, which is the leading order in a widely-used symmetry-preserving DSE truncation scheme \cite{Munczek:1994zz,Bender:1996bb}.  The second model is a truncation of the Ball-Chiu \emph{Ansatz} \cite{Ball:1980ay}.  It is far from the most general form \cite{Kizilersu:2009kg,Bashir:2011dp} but, in circumstances we expose, it produces some qualitative changes in Eq.\,\eqref{gendseN} and thus serves to highlight the impact of a dressed vertex on the number and nature of solutions to the gap equation. 

In general one can construct the Bethe-Salpeter kernel, $\Lambda_{J\nu}(q,\ell;K)$, associated with any $\Gamma_\nu(q,p)$ using the formulae in Ref.\,\cite{Chang:2011ei}.  Herein, however, $\Lambda_{J\nu}(q,\ell;K)$ is omitted for reasons we now explain.  This term is identically zero in rainbow-ladder truncation \cite{Chang:2009zb}; i.e., with Eq.\,\eqref{CalVRL}.  Hence the omission need only be discussed in connection with Eq.\,\eqref{CalV1BC}.
Firstly, Eq.\,\eqref{CalV1BC} has the same Dirac structure as Eq.\,\eqref{CalVRL} and hence the associated $\Lambda_{J\nu}(q,\ell;K)$ cannot realistically have a large effect on masses obtained via the Bethe-Salpeter equation since it does vanish identically using Eq.\,\eqref{CalVRL}.
More generally, we use the Bethe-Salpeter equation primarily in order to gauge stability of solutions to the gap equation.  A solution is stable if, and only if, both the scalar and pseudoscalar mass-squared values are positive semi-definite when computed using that solution.  In the scalar channel the omission of $\Lambda_{S\nu}(q,\ell;K)$ suppresses repulsion and hence produces a lower bound on the absolute value of the mass-squared \cite{Roberts:1996jx}.
In the pseudoscalar channel, on the other hand, the diagrams represented by $\Lambda_{P\nu}(q,\ell;K)$ largely cancel amongst themselves in the neighbourhood of the chiral limit, so this term has a negligible impact on the mass-squared within this domain \cite{Bender:2002as,Bhagwat:2004hn}.
Hence the omission of $\Lambda_{J\nu}(q,\ell;K)$ cannot materially affect a study of stability.


It is worth remarking that, irrespective of the remarks just made, all kernels constructed using Eqs.\,\eqref{CalG}, \eqref{CalV} preserve the one-loop renormalisation-group behavior of QCD in the gap and Bethe-Salpeter equations.  In the infrared, on the other hand, there are differences between Eq.\,\eqref{CalGMT} and \eqref{CalGQC}.  They are detailed in Ref.\,\cite{Qin:2011dd}; and notable amongst them is the fact that interactions constructed from Eq.\,\eqref{CalGQC} possess an infrared momentum-dependence that is consonant with modern DSE- and lattice-QCD results, whereas those produced by Eq.\,\eqref{CalGMT} violate this constraint.  Whilst this does not appear to impair the utility of Eq.\,\eqref{CalGMT} in connection with properties of ground state vector and flavour-nonsinglet pseudoscalar mesons, and the nucleon and $\Delta$, it does markedly affect its predictions for quantities more sensitive to the infrared behaviour of the interaction, such as the properties of excited states \cite{Qin:2011dd,Qin:2011xq} and, as we shall see, the location of phase boundaries.


\section{Results and Discussion}
\label{sec:results}
\subsection{Solutions of the Quark DSE}
In solving the gap equation we have two control parameters upon which the existence and number of solutions will depend: current-quark mass, $\hat m$; and interaction strength, which will hereafter be characterised by the dimensionless number\footnote{\label{ft:one} For reference, in rainbow-ladder truncation the domain of reasonable interaction strengths; i.e., strengths with which one can hope to describe hadron phenomena, is $3 \lesssim {\cal I}\lesssim 13$ for Eq.\,\eqref{CalGMT} and $2 \lesssim {\cal I}\lesssim 8$ for Eq.\,\eqref{CalGQC}.} ${\cal I} = D/\omega^2$.
That DCSB is a possibility for $\hat m=0$ and ${\cal I}>{\cal I}^c$, where ${\cal I}^c$ is some critical value, guarantees that the gap equation does admit more than one solution: ${\cal I}^c$ is a $\hat m=0$ bifurcation point \cite{Atkinson:1986aw,Cheng:1994sd}.  With the existence of furcation points assured, one must anticipate that the straightforward iteration procedure used commonly to solve the gap equation will be inadequate to the task of locating all solutions and tracking their evolution as $\{\hat m,{\cal I}\}$ are varied.  In contrast, the homotopy continuation method, summarised in App.\,\ref{app:A1}, is well suited to this challenge.

\subsubsection{Influence of interaction strength}
\label{sec:InfluenceI}

\begin{figure}[t]
\includegraphics[width=1.0\linewidth]{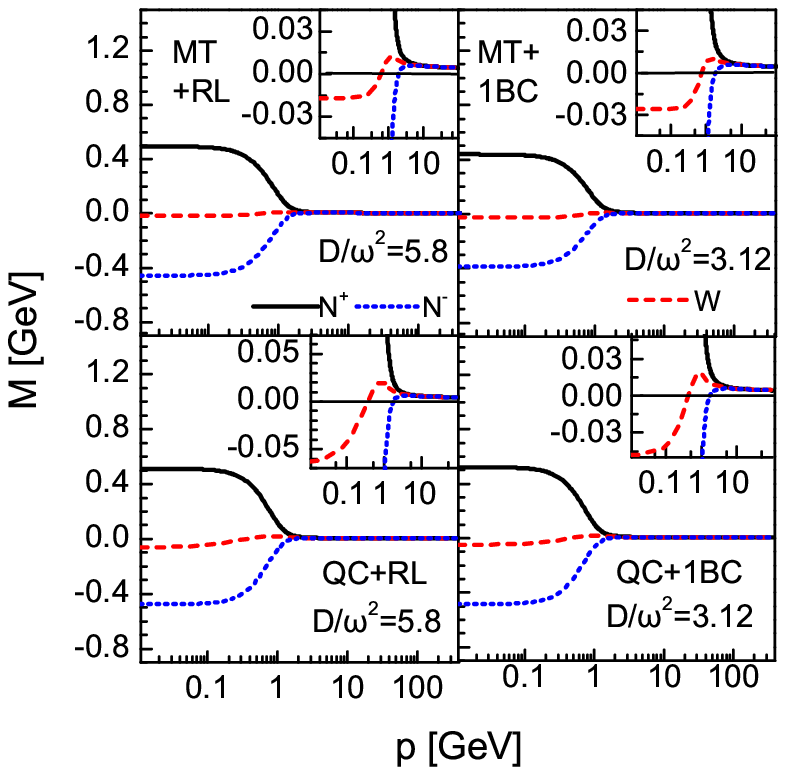}
\caption{Solutions of the gap equation obtained using $m^\zeta = 5\,$MeV with ${\cal I}=5.8$ for the rainbow-ladder vertex, Eq.\,\protect\eqref{CalVRL}, or ${\cal I}=3.1$ for the 1BC vertex, Eq.\,\protect\eqref{CalV1BC}.
Upper panels: interaction of Eqs.\,\protect\eqref{alphatilde}, \protect\eqref{CalGMT}, \protect\eqref{CalV}; lower panels: Eqs.\,\protect\eqref{alphatilde}, \protect\eqref{CalGQC}, \protect\eqref{CalV}.
All panels: solid curve, positive Nambu solution -- $N^+$; dashed, Wigner solution; and dotted, negative Nambu solution -- $N^-$.  The insets highlight the infrared behaviour of the Wigner and $N^-$ solutions, in particular their single zero.
\label{fig:DynamicalMass-All}}
\end{figure}

To illustrate the point and establish a context we solved the gap equation with the four interaction kernels described above.  The resulting mass functions are depicted in Fig.\,\ref{fig:DynamicalMass-All}.  With the listed parameters, the gap equation possesses three distinct solutions, as elucidated in Refs.\,\cite{Zong:2004nm,Chang:2006bm,Williams:2006vva}.  The figure displays one novelty, however: viz., both interactions support three nontrivial solutions with a dressed-vertex.  This dressing, albeit apparently simple, does qualitatively change the gap equation, as we will explain below.

\begin{figure}[t]
\includegraphics[width=1.0\linewidth]{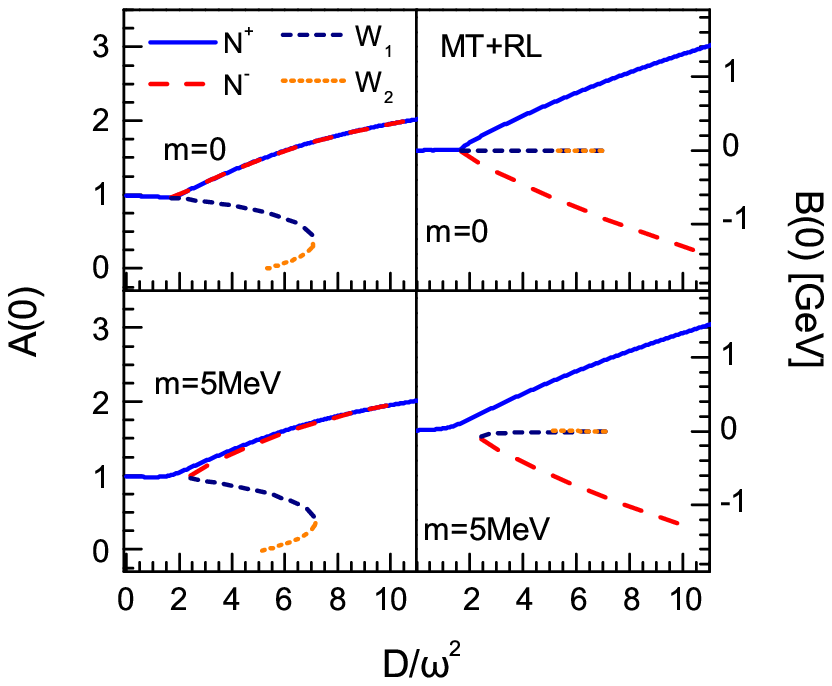}
\vspace*{-6ex}

\includegraphics[width=1.0\linewidth]{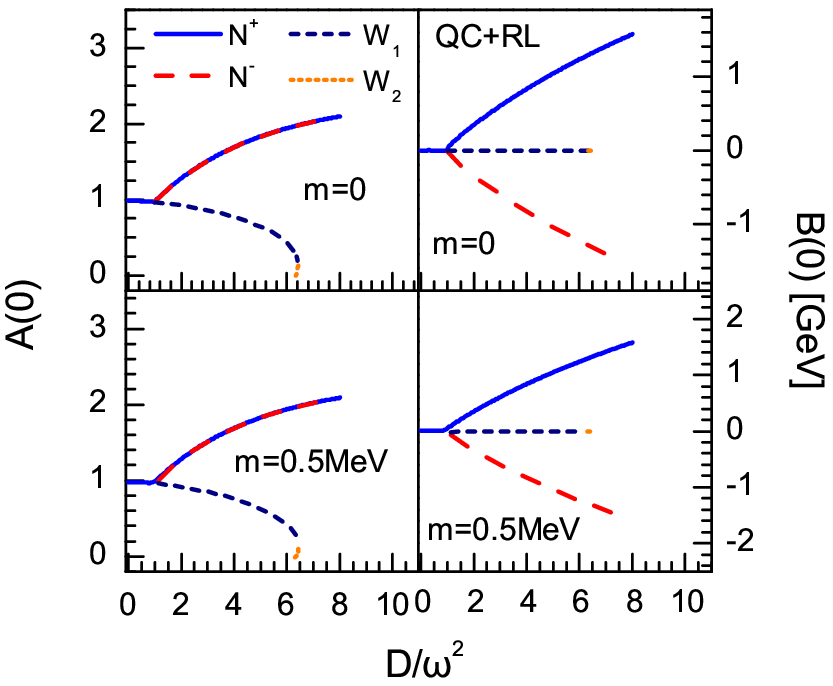}
\caption{
${\cal I}=D/\omega^2$-dependence of $A(0)$ and $B(0)$ obtained using the models specified by Eqs.\,\protect\eqref{alphatilde}, \protect\eqref{CalG}, \protect\eqref{CalVRL}:
upper grouping -- Eq.\,\protect\eqref{CalGMT}; and lower grouping -- Eq.\,\protect\eqref{CalGQC}.
The upper two panels in each grouping were computed in the chiral limit, whereas the lower two panels were obtained with $m^\zeta = 5\,$MeV or $m^\zeta = 0.5\,$MeV, respectively.
All panels: solid curve -- $N^+$ solution; long-dashed -- $N^-$; short-dashed -- regular Wigner solution; and dotted -- a second Wigner-type solution.  Naturally, in the chiral limit $A_{N^+} = A_{N^-}$; and deviations from this identity are almost imperceptible for $\hat m$ nonzero but small.
\label{fig:AM0-D-MTBare}}
\end{figure}

In order to determine the solution set of the gap equation, which, recall, is a pair of coupled, nonlinear integral equations for two functions, we solved Eq.\,\eqref{gendseN} on a large domain of $\{\hat m,{\cal I}\} \in \mathbb{R}^2$.  The values and parameter-dependence of the computed quantities $A(0)$ and $B(0)$ are useful in characterising the solutions.  Some of this information is portrayed in Fig.\,\ref{fig:AM0-D-MTBare}.  It is immediately apparent that, in the chiral limit, three critical points exist within the domain displayed.

The first is a trifurcation point.  For ${\cal I} < {\cal I}^c_1$, the magnitude of which depends on the form chosen from Eq.\,\eqref{CalG}, only the long known chiral-symmetry-preserving (Wigner) solution is present, which we hereafter denote $W$ or $W_1$.  At ${\cal I} = {\cal I}^c_1$, however, two new solutions appear.  These are the normal DCSB (Nambu) solutions, described above, which we henceforth denote $N^+$ or $N_1^+$ and $N^-$ or $N_1^-$, respectively.

\begin{figure}[t]
\includegraphics[width=1.0\linewidth]{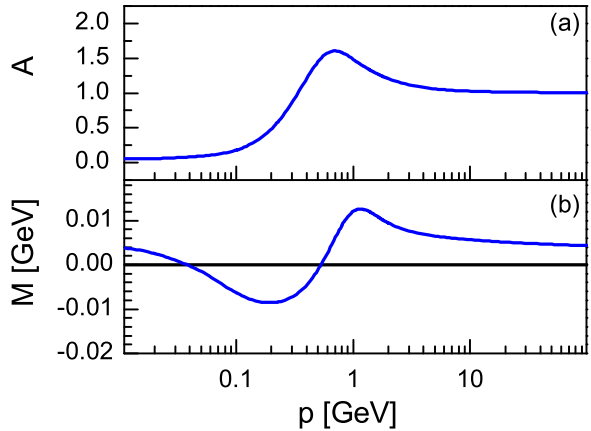}
\caption{Momentum dependence of the $W_2$ solution obtained using Eqs.\,\protect\eqref{alphatilde}, \protect\eqref{CalGMT}, \protect\eqref{CalVRL} with ${\cal I} = 5.8$ and $m^\zeta =5\,$MeV.  N.B.\ $M(p^2) \equiv 0$ for $\hat m = 0$. \label{fig:MemDepd-W2-MTBare}}
\end{figure}

The second critical point is associated with the appearance of a novel solution first observed in Ref.\,\cite{Williams:2006vva}.  At ${\cal I} > {\cal I}^c_2$, again interaction-dependent, a second Wigner-like solution, $W_2$, appears: whilst $B_{W_2}(p^2) \equiv 0$, $A_{W_2}(p^2)$ is nontrivial and $A_{W_2}(p^2)\neq A_{W_1}(p^2)$.  The momentum-dependence of the $W_2$ solution is depicted in Fig.\,\ref{fig:MemDepd-W2-MTBare} for small but nonzero current-quark mass. Plainly, the mass function has two zeros.

\begin{figure}[t]
\includegraphics[width=1.0\linewidth]{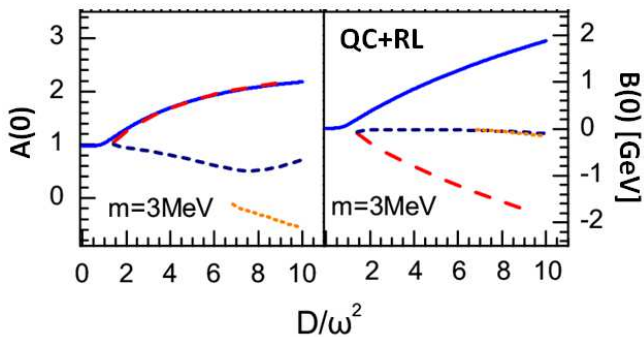}
\caption{
${\cal I}=D/\omega^2$-dependence of $A(0)$ and $B(0)$ obtained using the model specified by Eqs.\,\protect\eqref{alphatilde}, \protect\eqref{CalGQC}, \protect\eqref{CalVRL} and with $m^\zeta = 3\,$MeV.
Solid curve -- $N^+$ solution; long-dashed -- $N^-$; short-dashed -- regular Wigner solution; and dotted -- a second Wigner-type solution.
\label{fig:four}}
\end{figure}

A third critical point ${\cal I} = {\cal I}^c_3$ locates an interaction strength at which the $W_1$ and $W_2$ solutions merge and beyond which they disappear.  In the context of App.\,\ref{app:A1}, it is a turning point.  This is a novel result; for whilst the equation for $B(p^2)$ always admits the $B\equiv 0$ solution, the existence of ${\cal I}^c_3$ indicates that if the coupling strength is strong enough, then the equation for $A$ does not possess a solution.  We have thus exposed two chiral-limit examples of gap equations that only support a nonperturbative chiral symmetry preserving solution on a bounded domain of interaction strength.  Hence, one cannot in future assume that a gap equation will always admit a fully self-consistent Wigner solution at strong coupling.

\begin{figure}[t]
\includegraphics[width=1.0\linewidth]{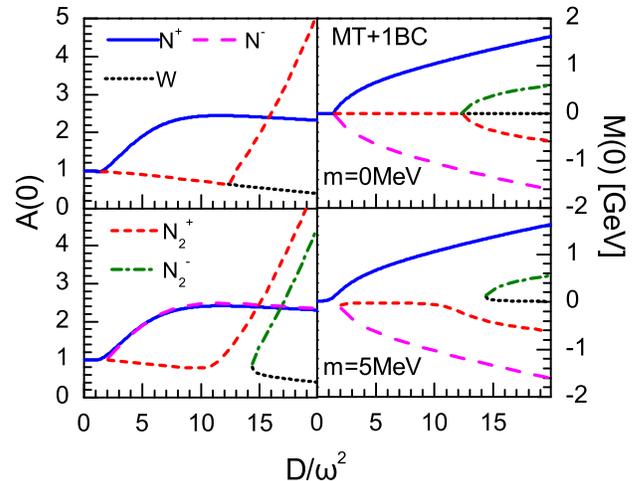}
\vspace*{-6ex}

\includegraphics[width=1.0\linewidth]{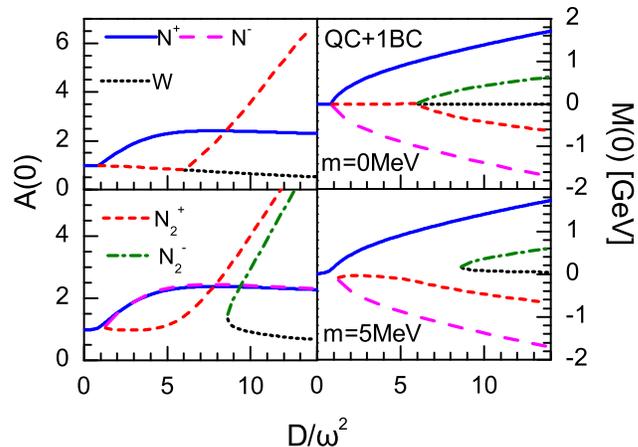}
\caption{
${\cal I}=D/\omega^2$-dependence of $A(0)$ and $M(0)$ obtained using the models specified by Eqs.\,\protect\eqref{alphatilde}, \protect\eqref{CalG}, \protect\eqref{CalV1BC}:
upper grouping -- Eq.\,\protect\eqref{CalGMT}; and lower grouping -- Eq.\,\protect\eqref{CalGQC}.
The upper two panels in each grouping were computed in the chiral limit, whereas the lower two panels were obtained with $m^\zeta = 5\,$MeV.
All panels: solid curve -- $N^+$ solution; long-dashed -- $N^-$; short-dashed -- $N_2^+$; dot-dashed -- $N_2^-$; and dotted -- Wigner solution.
Naturally, in the chiral limit $A_{N_i^+} = A_{N_i^-}$, $i=1,2$; and deviations from these identities are almost imperceptible for $\hat m$ nonzero but small.
\label{fig:AM0-D-MT1BC}}
\end{figure}

The picture changes somewhat at nonzero current-quark mass.  Of particular note: whilst the $N_1^+$ solution is always present, the $W_1$ and $N_1^-$ solutions exist only on a domain ${\cal I} \geq {\cal I}^{c_m}_1 > {\cal I}^c_1$: ${\cal I} = {\cal I}^{c_m}_1$ is a turning point.  Figure~\ref{fig:DynamicalMass-All} shows that in this case $M_{W}(p^2)$ is nonzero and both $M_{W}(p^2)$, $M_{N^-}(p^2)$ possess a zero.  Also striking is the sensitivity of the $W_1$, $W_2$ solutions to the infrared behaviour of the interaction, which is evident via comparison of Figs.\,\ref{fig:MemDepd-W2-MTBare} and \ref{fig:four}.  In the neighbourhood of the chiral limit, both $W_1$ and $W_2$ are absent for ${\cal I} > {\cal I}^c_3$, irrespective of the interaction.  Note, however, that one must be very close to $\hat m=0$ for this to be true when the interaction is constructed using Eq.\,\eqref{CalGQC}.  In this case there is a current-quark mass, $\hat m_{c_3}^{\rm QC}$, above which both $W_1$ and $W_2$ survive and evolve smoothly on ${\cal I} \geq {\cal I}^{c_m}_3 > {\cal I}^c_3$ (see Fig.\,\ref{fig:four}).  This is actually also true when the interaction is constructed using Eq.\,\eqref{CalGMT} but $\hat m_{c_3}^{\rm MT}/\hat m_{c_3}^{\rm QC}\gtrsim 10$.

The preceding few paragraphs described properties of solutions obtained with gap equation interaction kernels built in the rainbow truncation; i.e., using Eq.\,\eqref{CalVRL}.  Results obtained with the modestly dressed vertex in Eq.\,\eqref{CalV1BC} are displayed in Fig.\,\ref{fig:AM0-D-MT1BC}.  A comparison of Fig.\,\ref{fig:AM0-D-MT1BC} with Figs.\,\ref{fig:AM0-D-MTBare},\,\ref{fig:four} reveals a significant difference; viz., using Eq.\,\eqref{CalV1BC} there is only ever one Wigner-type solution.  A little algebra explains why: using Eq.\,\eqref{CalV1BC} the equation for $A(p^2)$ derived from Eq.\,\eqref{gendseN} is actually a linear equation for $\sigma_V(p^2)$ when $B\equiv 0$; and linear equations have at most one solution.  (N.B.\ This is also true when the ``2BC'' vertex is used; i.e., the vertex obtained from that in Ref.\,\cite{Ball:1980ay} by dropping only the Dirac-scalar term.)

\begin{figure}[t]
\includegraphics[width=1.0\linewidth]{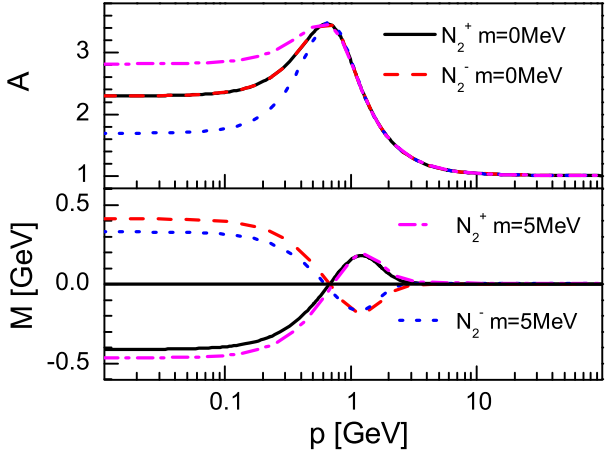}
\caption{
Momentum dependence of the $N_2^{\pm}$ solutions obtained using Eqs.\,\protect\eqref{alphatilde}, \protect\eqref{CalGMT}, \protect\eqref{CalV1BC} with ${\cal I} = 15.6$.
In both panels: solid curve -- $N_2^+$, chiral limit; long-dashed -- $N_2^-$, chiral limit; dot-dashed -- $N_2^+$, $m^\zeta=5\,$MeV; and dotted -- $N_2^-$, $m^\zeta=5\,$MeV.
\label{fig:AMp-MT1BC}}
\end{figure}

On the other hand, with increasing interaction strength, the number of Nambu-like solutions also grows.  The solutions we've labelled as $N_2^\pm$ each possess a single zero in the chiral limit, irrespective of the choice made in Eq.\,\eqref{CalG}; and the $N_2^-$ solution has two zeros when $\hat m^\zeta > 0$ as a result of being required to equal this positive mass at the renormalisation point (see Fig.\,\ref{fig:AMp-MT1BC}).
The momentum-dependence of the new Nambu solutions becomes quite complicated as the interaction strength reaches large values.  Notwithstanding this, there is always a value of the interaction strength above which these solutions exhibit the hallmarks of the normal Nambu solutions; i.e., in the chiral limit they are nonzero mirror image pairs, and for small nonzero current-quark mass the members of the pair have commensurate magnitudes.

We define Nambu-like to mean solutions with a nonzero mass function in the chiral limit.  However, our nomenclature is not without ambiguity.  For example, on $ 1 \lesssim {\cal I} \lesssim 6$ the $N_2^+$ solution has properties characteristic of a Wigner solution: the mass function is zero and it trifurcates from the regular Nambu solutions at the lower boundary of this domain, evolving thereafter within the domain as a chirally symmetric solution.  At the upper end, however, it trifurcates instead as the partner to the $N_2^-$ solution and maintains that DCSB trajectory.  It is finally for this reason that we label it a Nambu-like solution.  This pattern might repeat again with increasing ${\cal I}$, so that the solution we have labelled as Wigner-like is, in fact, the $N_3^+$ solution, which will, in turn, trifurcate to form the DCSB partner of the $N_3^-$ solution, leaving either a true Wigner solution or a $N_4^+$ solution, if the pattern is interminable.  Whilst this is of academic interest it is not physically relevant since the values of ${\cal I}$ involved far exceed the upper bound on values which are capable of producing an efficacious hadron physics phenomenology.

\subsubsection{Influence of current-quark mass}
\label{sec:Influencem}

\begin{figure}[t]
\vspace*{-4ex}

\includegraphics[width=1.0\linewidth]{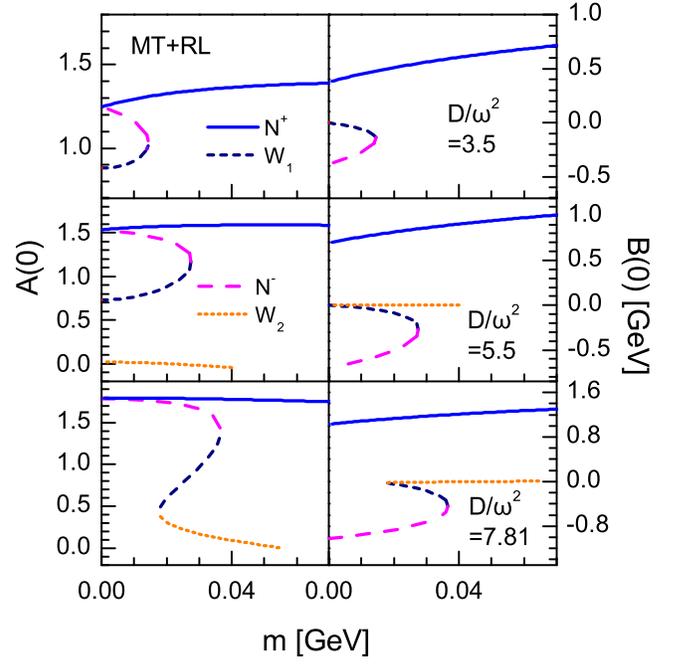}
\caption{
Current-quark-mass-dependence of $A(0)$ and $B(0)$ obtained using the model specified by Eqs.\,\protect\eqref{alphatilde}, \protect\eqref{CalGMT}, \protect\eqref{CalVRL}:
upper pair -- ${\cal I}=3.5$; middle pair -- ${\cal I}=5.5$; and bottom pair -- ${\cal I}=7.8$.
All panels: solid curve -- $N^+$ solution; short-dashed -- $W_1$; long-dashed -- $N^-$; and dotted -- $W_2$.
\label{fig:AM0-m-MTBare}}
\end{figure}

\begin{figure}[t]
\vspace*{-4ex}

\includegraphics[width=1.0\linewidth]{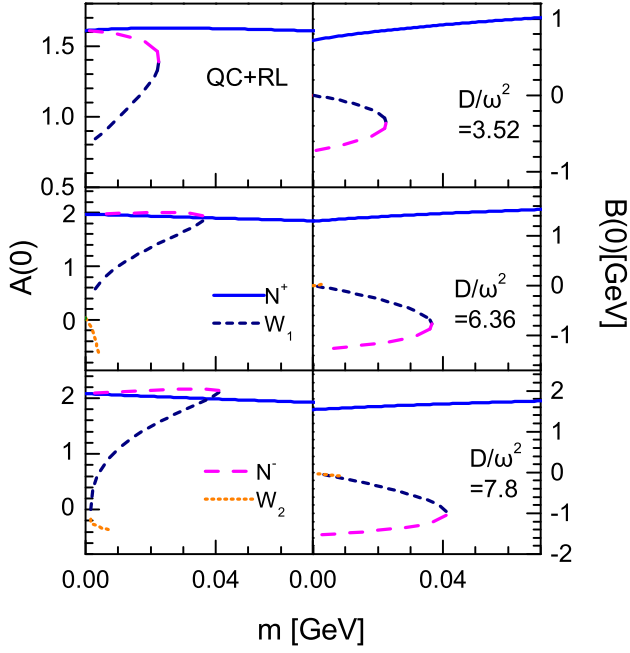}
\caption{
Current-quark-mass-dependence of $A(0)$,\,$B(0)$ obtained using the model specified by Eqs.\,\protect\eqref{alphatilde}, \protect\eqref{CalGQC}, \protect\eqref{CalVRL}:
upper pair -- ${\cal I}=3.5$; middle pair -- ${\cal I}=6.4$; and bottom pair -- ${\cal I}=7.8$.
All panels: solid curve -- $N^+$ solution; short-dashed -- $W_1$; long-dashed -- $N^-$; and dotted -- $W_2$.
\label{fig:AM0-m-ICBare}}
\end{figure}

It will already be plain from Sec.\,\ref{sec:InfluenceI} that the nature and number of solutions to the gap equation also depend on the current-quark mass.  This is emphasised by Figs.\,\ref{fig:AM0-m-MTBare}, \ref{fig:AM0-m-ICBare}, which show that the simultaneous existence of distinct solutions depends sensitively on the location in $\mathbb{R}^2$ of the control parameters $\{\hat m,{\cal I}\}$.  The Wigner solutions are again a good example.  There are points $\{\hat m,{\cal I}\} \in \mathbb{R}^2$ at which $A_{W_2}(0)=0$ and only with sufficiently large interaction strength is there a clear relationship between the $W_1$ and $W_2$ solutions.   At such strengths, however, the Wigner solutions do not exist in a sizeable connected domain containing the chiral limit.  (These points were mentioned earlier, in connection with Figs.\,\ref{fig:AM0-D-MTBare},\,\ref{fig:four}.)

\begin{figure}[t]
\includegraphics[width=1.0\linewidth]{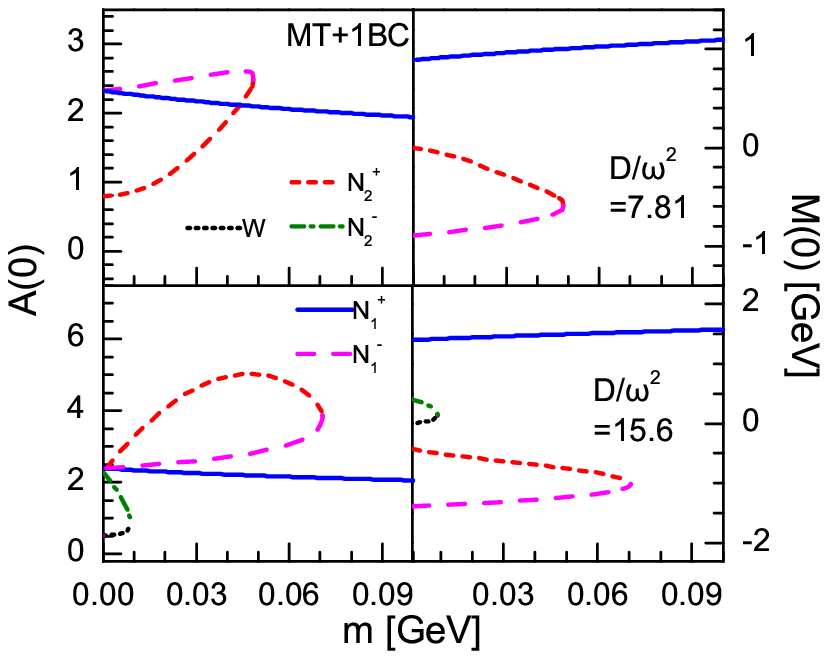}
\vspace*{-4ex}

\includegraphics[width=1.0\linewidth]{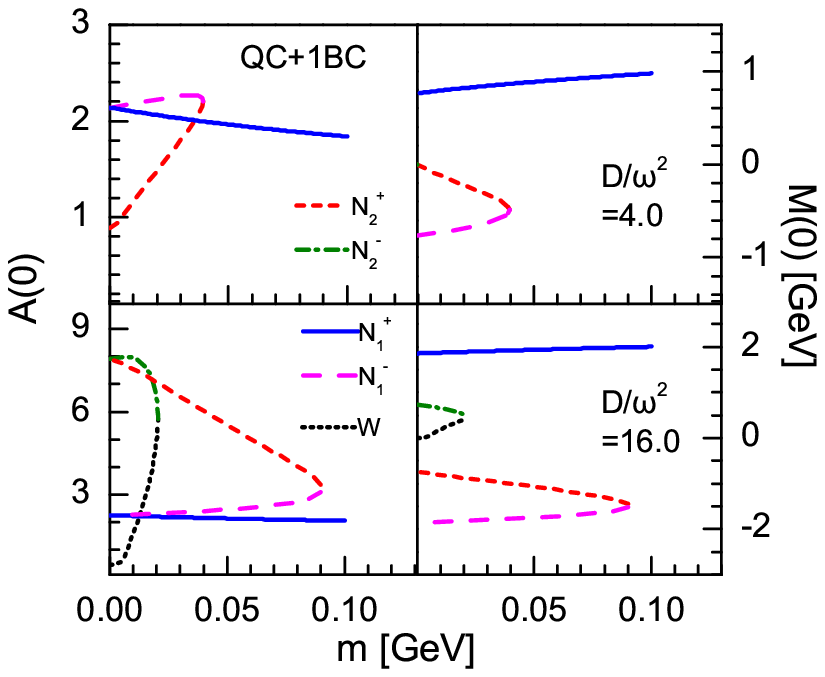}

\caption{
Current-quark-mass-dependence of $A(0)$,\,$M(0)$ obtained using the models specified by Eqs.\,\protect\eqref{alphatilde}, \protect\eqref{CalG}, \protect\eqref{CalV1BC}:
upper grouping -- Eq.\,\protect\eqref{CalGMT}, with ${\cal I}= 7.8$ (top row) and ${\cal I}= 15.6$ (bottom row); and lower grouping -- Eq.\,\protect\eqref{CalGQC}, with ${\cal I}=4$ (top row) and ${\cal I}= 16$ (bottom row).
All panels: solid curve -- $N_1^+$ solution; long-dashed -- $N_1^-$; short-dashed -- $N_2^+$; dot-dashed -- $N_2^-$; and dotted -- Wigner solution.
\label{fig:AM0-m-MT1BC}}
\end{figure}

The influence on the solutions of dressing the quark-gluon vertex is illustrated in Fig.\,\ref{fig:AM0-m-MT1BC}.  In important respects, the picture is simpler in this case.  As usual, the regular Nambu solution is distinct but all other solutions can be considered to appear in pairs, something we noted earlier in connection with Fig.\,\ref{fig:AM0-D-MT1BC}.  One simple observation is important and supported by the figure; viz., at fixed interaction strength the number of solution pairs decrements uniformly as the current-quark mass passes discrete critical values until only the regular Nambu solution exists.

\subsection{Solution Set Domains}
The results described in Secs.\,\ref{sec:InfluenceI},\,\ref{sec:Influencem} indicate that while choosing between Eqs.\,\eqref{CalGMT} and \eqref{CalGQC} produces quantitative changes, both interactions produce a qualitatively similar solution set.  Dressing the quark-gluon vertex, however, produces qualitative changes as well.  This is consistent with recent studies that have highlighted the impact of vertex dressing on hadron phenomena \cite{Chang:2009zb,Chang:2010hb,Chang:2011ei} and can be elucidated in the present context by charting the solution domains.

\begin{figure}[t]
\includegraphics[width=1.0\linewidth]{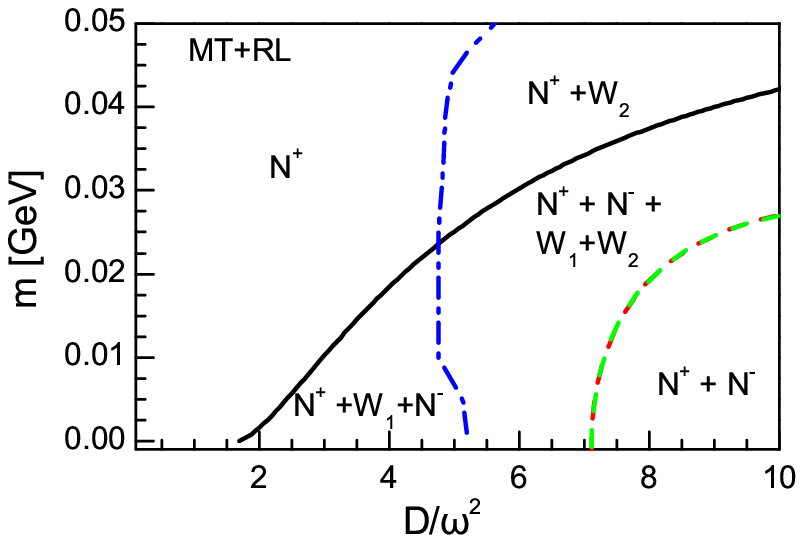}
\vspace*{-4ex}

\includegraphics[width=1.0\linewidth]{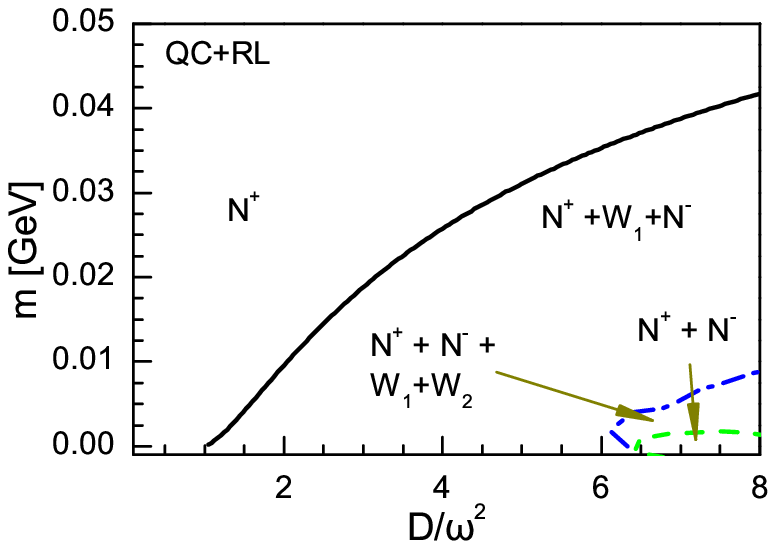}
\caption{Chart of gap equation solution domains in the $\{ m^\zeta , {\cal I} = D/\omega^2 \}$-plane obtained with the models specified by Eqs.\,\protect\eqref{alphatilde}, \protect\eqref{CalG}, \protect\eqref{CalVRL}:
top panel -- Eq.\,\protect\eqref{CalGMT}; and bottom panel -- Eq.\,\protect\eqref{CalGQC}.
In both panels the annotations within the bounded regions indicate which solutions are found.
\label{fig:PhaseDiagram-Bare}}
\end{figure}

\begin{figure}[t]
\includegraphics[width=1.0\linewidth]{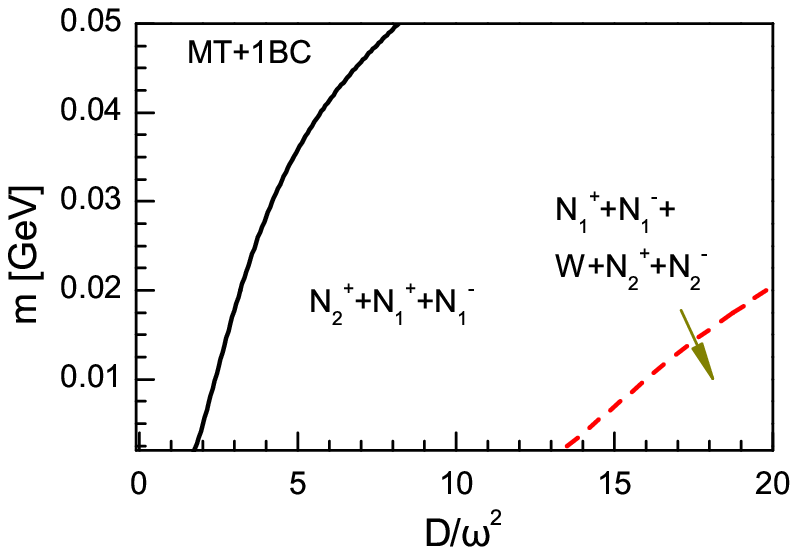}
\vspace*{-4ex}

\includegraphics[width=1.0\linewidth]{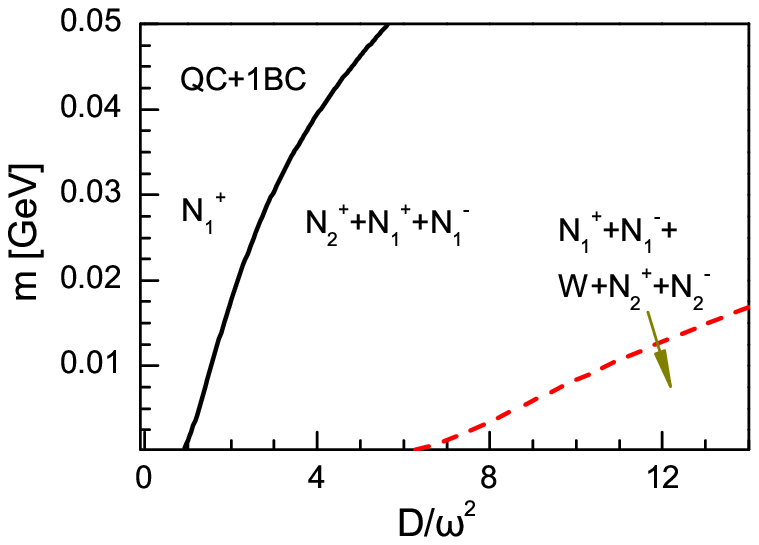}
\caption{Chart of gap equation solution domains in the $\{ m^\zeta , {\cal I} = D/\omega^2 \}$-plane obtained with the models specified by Eqs.\,\protect\eqref{alphatilde}, \protect\eqref{CalG}, \protect\eqref{CalV1BC}:
top panel -- Eq.\,\protect\eqref{CalGMT}; and bottom panel -- Eq.\,\protect\eqref{CalGQC}.
In both panels the annotations within the bounded regions indicate which solutions are found.
\label{fig:eleven}}
\end{figure}

In Fig.\,\ref{fig:PhaseDiagram-Bare} we display the gap equation solution domains for the rainbow truncation, whereas those for the 1BC truncation are depicted in Fig.\,\ref{fig:eleven}.   We remark that these figures were computed with $\omega=0.4\,$GeV for Eq.\,\eqref{CalGMT} and $\omega=0.5\,$GeV for Eq.\,\eqref{CalGQC}.  However, we did vary these parameters within the domains described in connection with Eqs.\,\eqref{CalG} and found that there is little variation.

It is clear from the figures that little of interest is possible until the interaction strength is sufficient to support nonperturbative solutions to the gap equation.  Thereafter, however, the rainbow truncation can produce a novel Wigner-like solution, $W_2$, whose momentum dependence is typified by Fig.\,\ref{fig:MemDepd-W2-MTBare}, but this is particular to that truncation.  The studies with a modestly dressed vertex show a simpler, regular pattern.  We have already noted that the gap equation's solution set is quite complicated at very large interaction strengths.  However, such strengths far exceed the upper bound on values which are capable of describing hadron observables (see footnote~\ref{ft:one}) and hence we do not describe them herein.

Let us imagine for the moment that QCD's gap equation possesses a kernel whose solution set is one of the complicated domains.  It may be argued that the different solutions within a domain represent competing phases; and should they exist simultaneously, then the computation of hadron properties would become a complicated affair.  Moreover, their existence would likely be reflected in the properties of hadrons.  Since the vast body of DSE-based hadron phenomenology does not show any sign that this is the case \cite{Chang:2011vu,Bashir:2012fs,Roberts:2012sv}, there must be an egress.

\subsection{Phase Stability}
\begin{figure}
\includegraphics[width=1.0\linewidth]{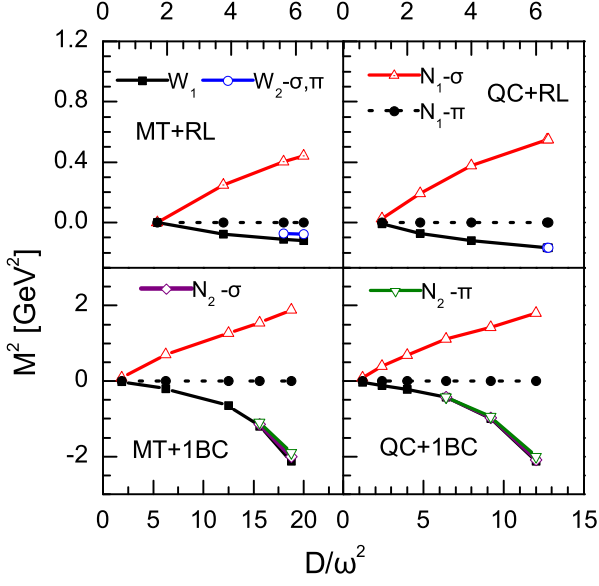}
\caption{${\cal I}=D/\omega^2$-dependence of $m_{\pi}^{2}$ an $m_{\sigma}^{2}$ obtained via the Bethe-Salpeter equations in the chiral limit.  The panels depict results obtained with different kernels; namely, Eq.\,\protect\eqref{alphatilde} and: upper-left -- Eqs.\,\protect\eqref{CalGMT}, \protect\eqref{CalVRL}; upper-right -- Eqs.\,\protect\eqref{CalGQC}, \protect\eqref{CalVRL}; lower-left -- Eqs.\,\protect\eqref{CalGMT}, \protect\eqref{CalV1BC}; and lower right Eqs.\,\protect\eqref{CalGQC}, \protect\eqref{CalV1BC}.
All panels: filled squares -- $m_\pi^2=m_\sigma^2$ along the $W_1$ solution trajectory; open circles -- $m_\pi^2=m_\sigma^2$ along $W_2$; open up-triangles -- $m_\sigma^2$ along $N_1^+$; filled circles -- $m_\pi^2$ along $N_1^+$; open diamonds -- $m_\sigma^2$ along $N_2^+$; and down-triangles -- $m_\pi^2$ along $N_2^+$.
\label{fig:mpis-D}}
\end{figure}

\begin{figure}
\includegraphics[width=0.89\linewidth]{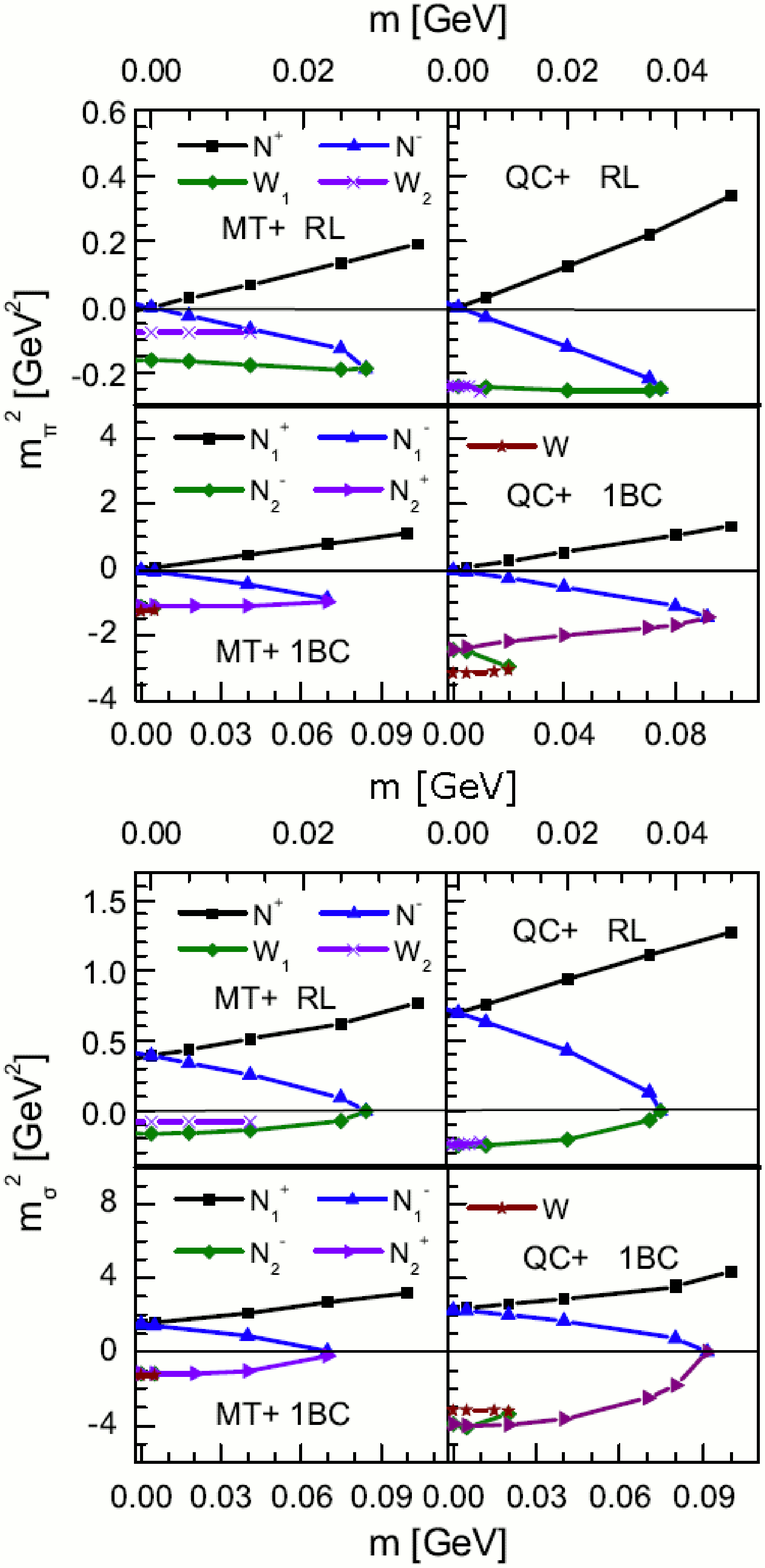}

\caption{
Current-quark-mass-dependence of $m_\pi^2$ (upper grouping) and $m_\sigma^2$ (lower grouping) obtained with different kernels; namely, Eq.\,\protect\eqref{alphatilde} and, within each grouping: upper-left -- Eqs.\,\protect\eqref{CalGMT}, \protect\eqref{CalVRL}, ${\cal I}=5.5$; upper-right -- Eqs.\,\protect\eqref{CalGQC}, \protect\eqref{CalVRL}, ${\cal I}=6.4$; lower-left -- Eqs.\,\protect\eqref{CalGMT}, \protect\eqref{CalV1BC}, ${\cal I}=15.6$; and lower right Eqs.\,\protect\eqref{CalGQC}, \protect\eqref{CalV1BC}, ${\cal I}=16.0$.
Upper row of each grouping:
squares -- $m_J^2$, $J=\sigma,\pi$ along the $N_1^+$ solution trajectory;
up-triangles -- $m_J^2$ along $N_1^-$;
diamonds -- $m_J^2$ along $W_1$
and crosses -- $m_J^2$ along $W_2$.
Lower row of each grouping:
squares -- $m_J^2$ along $N_1^+$;
up-triangles -- $m_J^2$ along $N_1^-$;
diamonds -- $m_J^2$ along $N_2^-$;
right-triangles -- $m_J^2$ along $N_2^+$;
and stars -- $m_J^2=m_\sigma^2$ along the sole Wigner trajectory.
\label{fig:mpis-m0}}
\end{figure}

Egress lies in the direction of phase stability.  One must consider which of the solutions within a given domain is stable against fluctuations.  Figures~\ref{fig:mpis-D} and \ref{fig:mpis-m0} contain the information necessary to address this question through the stability criterion introduced as the generalisation of Eqs.\,\eqref{d2Usp}.

A careful examination of Fig.\,\ref{fig:mpis-D} reveals that solutions of the inhomogeneous Bethe-Salpeter equations do not exhibit bound-state poles until ${\cal I}\geq {\cal I}_1^c$; i.e., until the interaction strength exceeds the critical value for DCSB.  (This is another example of the causal connection between confinement and DCSB in DSE models of QCD -- see, e.g., Sec.\,2 in Ref.\,\cite{Bashir:2012fs}.)   Amidst the solutions displayed beyond ${\cal I}= {\cal I}_1^c$, only the regular Nambu solution of the gap equation is stable: it produces the well-known DCSB case of a massless pseudoscalar meson accompanied by a massive scalar (point ``I'' in Fig.\,\ref{fig:MexHat}).  In the chiral limit the negative Nambu solution, its partner, produces the same results and is equally stable.  By the same token, the partners to the other displayed solutions are unstable.
Plainly, apart from the peculiarity of $W_2$, these results are qualitatively the same in all models considered.

Let us turn now to Fig.\,\ref{fig:mpis-m0}.  In all panels the interaction strength is large enough to induce DCSB at $\hat m = 0$; and it is abundantly clear that $m_\pi^2$ and $m_\sigma^2$ are only both positive semi-definite along the trajectory of the regular Nambu solution, $N_1^+$.  This is true on an unbounded domain of $\hat m>0$.  Unsurprisingly, given the qualitative connection between our stability criterion and a ``Mexican hat'' potential, the negative Nambu solution, $N_1^-$, is a saddle-point trajectory: $m_\sigma^2>0$ but $m_\pi^2<0$ (point ``S'' in Fig.\,\ref{fig:MexHat}).  This nature persists until the current-quark mass exceeds a critical value, whose magnitude is model-dependent but may be characterised as $m^\zeta_c \sim 0.06\pm0.03\,$GeV.  This was first observed in \cite{Chang:2006bm,Williams:2006vva}.  Inspection of the lower row in each grouping reveals the role of the $N_2^+$ trajectory as a surrogate Wigner solution within a connected, bounded domain of current-quark mass, just as was discussed in connection with Fig.\,\ref{fig:AM0-D-MT1BC}.  It is evident from Fig.\,\ref{fig:mpis-m0} that all other solutions correspond to unstable trajectories.

\begin{figure}[t]
\includegraphics[width=1.0\linewidth]{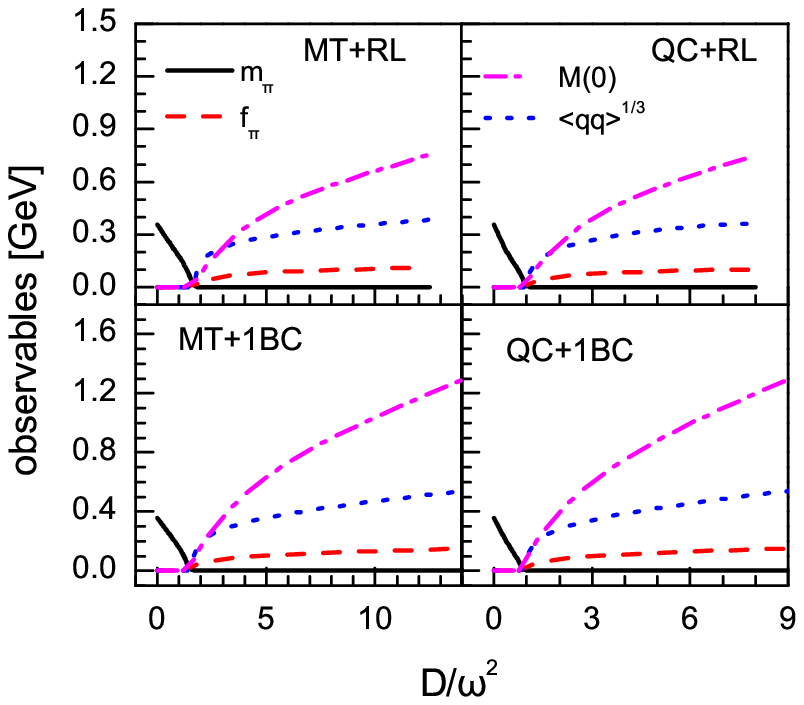}
\includegraphics[width=1.0\linewidth]{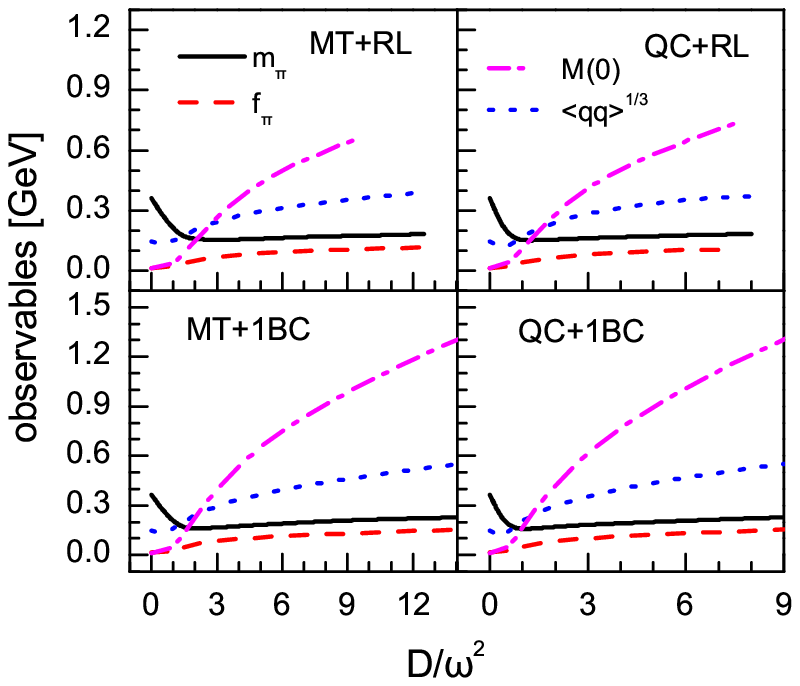}
\caption{
Calculated ${\cal I}=D/\omega^2$-dependence of a range of quantities that typify DCSB in hadron physics: solid curve -- $m_\pi$; long-dashed -- $f_\pi$; dot-dashed -- $M(0)$; and short-dashed -- in-pion condensate \protect\cite{Maris:1997tm,Brodsky:2010xf,Brodsky:2012ku}.
Upper grouping: chiral limit; and lower grouping $m^\zeta=5\,$MeV.
Within each grouping, the results were computed with Eq.\,\protect\eqref{alphatilde} and: upper-left -- Eqs.\,\protect\eqref{CalGMT}, \protect\eqref{CalVRL}; upper-right -- Eqs.\,\protect\eqref{CalGQC}, \protect\eqref{CalVRL}; lower-left -- Eqs.\,\protect\eqref{CalGMT}, \protect\eqref{CalV1BC}; and lower right Eqs.\,\protect\eqref{CalGQC}, \protect\eqref{CalV1BC}.
\label{fig14}
}
\end{figure}

In Fig.\,\ref{fig14} we depict the evolution with interaction strength of a range of quantities which typify DCSB in hadron physics.  Proceeding along the $N^+$ trajectory from large to small values of ${\cal I}=D/\omega^2$ in the upper grouping of panels, one sees behaviour typical of a Nambu to Wigner phase transition.  In the chiral limit, all order parameters for DCSB vanish at the critical interaction strength; and the sharp jump in $m_\pi$ shows coincident deconfinement.  These panels are qualitatively identical to Fig.\,3 in Ref.\,\cite{Maris:2000ig}, which shows the temperature dependence of these and related quantities through the temperature induced deconfinement and chiral symmetry restoration phase transitions.  Notably, here as there, $f_\pi$ vanishes because at ${\cal I}<{\cal I}_1^c$ the pseudoscalar correlation involving deconfined quarks possesses neither  pseudovector nor pseudotensor components; i.e., $F\equiv 0 \equiv G\equiv H$ in Eq.\,\eqref{GammaP}.

With the introduction of a small current-quark mass the transitions become a crossover, as evident in the lower grouping of panels in Fig.\,\ref{fig14}.  Nevertheless, deconfinement is still evident through the sharp rise in the trajectory associated with $m_\pi$ below ${\cal I}_1^c$, a domain whereupon the mass-scale determining the magnitude of $f_\pi$, $M(0)$ is seen to switch to the explicit chiral symmetry breaking current-quark mass.  These results are qualitatively identical to those in Fig.\,6 of Ref.\,\cite{Maris:2000ig}.

The gap equation has long been used as a tool for identifying field configurations that may optimally be employed in constructing a mean-field approximation, or improvements thereof, to a theory's generating functional.  Indeed, truncated gap equations are critical in the construction of effective actions for composite operators \cite{Haymaker:1990vm} and therefrom developing models for DCSB in hadron physics.  In this connection, the gap equation solutions have often been interpreted as candidate vacua, some of them energetically equivalent but distinct, related via global symmetry transformations, in the sense first described in connection with the Nambu--Jona-Lasinio model \cite{Nambu:1961tp,Nambu:2011zz}.  Some external agent, such as current-quark mass, then tips the balance in favour of one solution, which therefore provides the configuration around which a model Lagrangian is constructed to describe field fluctuations; e.g., Refs.\,\cite{Finger:1981gm,Cahill:1985mh,McKay:1987nx,Roberts:1994dr}.  Such models typically arrive at a potential which expresses features that are synonymous with those of the ``Mexican hat''.  These observations provide another context for our results; viz., the models possess far more candidate vacua than practitioners had usually imagined, with an hierarchical structure such that, within levels, mappings exist between those solutions related by a symmetry transformation.  Notwithstanding this, the standard Nambu solution of the gap equation is the only one that is stable in the presence of a nonzero current-quark mass.

\section{Remarks and Summary}
\label{sec:epilogue}
We argued, by way of examples, that models of QCD's gap equation will typically possess many solutions, a feature which owes to the nonlinearity of the equation.  Although simple vertex Ans\"atze were used, we judge that results obtained with the 1BC form exemplify the behavior one should expect with more realistic models.

The nature and number of the solutions is readily explained and understood.  Naturally, in the weak coupling limit, only the usual perturbative -- Wigner type -- solution is possible.  On the other hand, the number of chiral-limit solutions evolves with interaction strength, so that at large interaction strengths there are many solutions, with distinct pointwise behaviour, that express dynamical chiral symmetry breaking (DCSB).  To be clear: there are numerous DCSB solutions in addition to that which is usually labelled as the Nambu solution.  In response to increasing current-quark mass, however, the number of solutions decrements uniformly as particular thresholds are crossed until, above some value of the mass, only the regular Nambu solution remains.

The gap equation's nonperturbative solutions form a hierarchy.  In the chiral limit there is a solution within each level that preserves chiral symmetry but also a set of distinct DCSB solutions that are energetically equivalent and related by a symmetry transformation.  A symmetry transformation does not connect solutions in different levels, however, and nor are solutions in different levels degenerate.

In the context of composite operator effective actions, solutions of the gap equation play the role of candidate vacua in the sense that one, from amongst all those available, should be chosen as the ground state about which dynamical fields may fluctuate.  A stability criterion is necessary before such a choice can be made.  One is readily derived from a consideration of the scalar and pseudoscalar susceptibilities via their explanation of the ``Mexican hat'' potential and relationship to the fully-dressed propagators for composite correlations in the scalar and pseudoscalar channels.  Fortunately for hadron physics phenomenologies, when applied to the array of gap equation solutions, this stability test shows that for any nonzero current-quark mass only the regular Nambu solution of the gap equation is stable against perturbations.

\begin{acknowledgments}
We are grateful for input from and discussion with A.~Bashir, A.~Raya, K.~Raya and D.\,J.~Wilson.
This work was supported by:
the National Natural Science Foundation of China, under contract nos.~10935001, 11075052 and 11175004; the Major State Basic Research Development Program, under contract no.~G2007CB815000;
Forschungszentrum J\"ulich GmbH;
and
U.\,S.\ Department of Energy, Office of Nuclear Physics, contract no.~DE-AC02-06CH11357.

\end{acknowledgments}

\appendix

\section{Homotopy continuation method}
\label{app:A1}
In the context of nonlinear integral equations the homotopy continuation method \cite{Homotopy} enables one to do more than follow a single path to a solution: one can also, e.g., switch branches at simple furcation points.  The approach is therefore more powerful and discriminating than simple iteration to a solution.  We illustrate aspects of the method here using the gap equation as an illustrative example.

To proceed one first converts the integral equation into a matrix equation using a discretisation method.  The Chebyshev expansion scheme described in Ref.\,\cite{Bloch:1995dd} is efficient.  The gap equation may then be represented as follows:
\begin{eqnarray}
X_i&=&\left\{\begin{array}{lll}
A(p^2_i)&\ &0\le i<N/2\\
B(p^2_i)&\ &N/2\le i<N
\end{array}\right.,\\\nonumber
F_i({X_j})&=&0,\  \qquad \quad \; \; \textrm{where} \ \ 0\le i,j<N.
\end{eqnarray}

Suppose now that one has a control parameter, $\lambda \in R$, upon which the solution of the gap equation depends.  Herein $\lambda = m$, the current-quark mass, or $\lambda=D/\omega^2$, the interaction strength.  Given a value of $\lambda$, the gap equation can be understood as the identity
\begin{equation}
H(u)=0_N, \qquad u_i=X_i \,, u_{N}=\lambda\,,
\label{eq:H0}
\end{equation}
where H: $\mathbb{R}^{N+1} \rightarrow \mathbb{R}^{N}$ is a smooth mapping on some closed domain ${\cal D} \in \mathbb{R}^{N+1}$ and $0_N$ is the null-vector in $\mathbb{R}^{N}$.  The solutions of Eq.\,\eqref{eq:H0} are an inverse image of the null-vector.  Denoted $H^{-1}(0_N)$, in general this inverse image describes a collection of smooth curves in $\mathbb{R}^{N+1}$.  Importantly, so long as $\forall u\in {\cal D}$: ${\rm rk}(H'(u))=N$; i.e., the derivative has maximal rank throughout ${\cal D}$, then each one of these curves begins and ends on $\partial {\cal D}$, the boundary of ${\cal D}$, and no two intersect.

In order to elucidate we will return to interpreting $\lambda$ as a control parameter, in which case solutions of the gap equation depend parametrically on this variable: $X=X(\lambda)$.  In a typical gap equation study one may view the solution process as beginning with some small nonnegative value of $\lambda$, locating the zero; then repeating the zero finding steps as $\lambda$ is smoothly incremented.  With this in mind, suppose that at a given value of $\lambda=\lambda_1$ the gap equation has a solution $X^1=X(\lambda_1)$; i.e., $F(X^1;\lambda_1)=0_N$.  Suppose in addition that one has already obtained the solution on some domain $\lambda_0\leq \lambda <\lambda_1$; i.e., one knows $X(\lambda)$ on this domain, and
\begin{equation}
\label{detneqzero}
\lim_{\lambda\to\lambda_1}{\rm det}\frac{\partial F(X;\lambda)}{\partial X} \neq 0\,,
\end{equation}
then $X^1=X(\lambda_1)$ is readily obtained via straightforward iteration from $X(\lambda_1^-)$; viz., the solution at some nearby $\lambda_1^-<\lambda_1$.

On the other hand, suppose $X(\lambda_1)$ is a solution but
\begin{equation}
{\rm det}\frac{\partial F(X;\lambda_1)}{\partial X}\bigg|_{X=X^1} = 0\,.
\label{detFzero}
\end{equation}
At such a point $X^1\in \mathbb{R}^N$, either ${\rm rk}(H'(u))= N$ or ${\rm rk}(H'(u))\neq N$.  Consider the first possibility, which corresponds to the curves $H^{-1}(0_N)$ being smooth.  In this case
\begin{equation}
\lim_{\lambda\to\lambda_1}\left|\frac{d X}{d\lambda}\right| =\infty
\end{equation}
and $X$ locates a singular point of one of the curves generated by $H^{-1}(0_N)$.

There are still two possibilities: in the neighbourhood of $\lambda_1$ the surface $X(\lambda)$ may either be characterised as possessing the form of a straightened S-bend or exhibiting a turning point; i.e., bending back upon itself.  In the first case it might be difficult to obtain the solution at $\lambda_1$ by iteration but this straightforward procedure will converge at $\lambda_1^+ = \lambda +\epsilon$, where $\epsilon$ is a small positive number that may be determined empirically.  The solution at $\lambda_1$ is then bracketed and may be found by interpolation; and one can continue the iterative procedure on $\lambda>\lambda_1$.

The situation is different at a turning point, which, in the context of our study, locates the critical current-quark mass for the transition between a Nambu solution and a Wigner solution.  Iteration fails at a turning point.  In this case one may proceed as follows.  Suppose one has a solution at $\lambda_1^-$:
\begin{equation}
\begin{array}{l}
X(\lambda_1^-) = \{x_0,x_1,\ldots,x_{N-1}\},\; \\
F(\{x_0,x_1,\ldots,x_{N-1}\};\lambda_1^-)=0_N.
\end{array}
\end{equation}
Now shift $x_i\to \hat x_i = x_i+\delta$, with $\delta$ some small number and, typically, $i=N/2$; hold $\hat x_i$ fixed; and solve by iteration the problem
\begin{equation}
F( \{\lambda,x_0,\ldots,\not \!x_i, \ldots,x_n\};\hat x_i)=0_N .
\label{parameterchange}
\end{equation}
This represents an interchange of roles between the control parameter and one element of the solution vector.  That which was previously the control parameter now becomes part of a modified solution vector that is sought by iteration.  This simple method enables one to join and follow the trajectory of the second solution, which exists simultaneously on $\lambda < \lambda_1$ with that already obtained.  Once one is sufficiently far removed from $\lambda_1$ on this new trajectory, straightforward iteration can again be employed.

%

Return now to Eq.\,\eqref{detFzero} and consider the remaining possibility; in particular, ${\rm rk}(H'(u)) = N -1$.  In principle this could correspond to one of the curves  $H^{-1}(0_N)$ terminating within ${\cal D}$.  For the gap equation, however, this is impossible because it would indicate that there is some domain of parameter space in which the gap equation does not have a solution.  Hence for us this case represents a point at which two inverse images of the null vector intersect.  To be more explicit, we encounter this situation when incrementing $\lambda = D/\omega^2$ in the chiral limit to arrive at a trifurcation point, whereat a Wigner solution connects with a Nambu solution and its reflection.  At such a location both iteration and the role change method of Eq.\,\eqref{parameterchange} fail in the sense that neither enables the subsequent trajectory of all solutions to be followed.

To circumvent this difficulty we exploit the current-quark mass; i.e., we solve $H(u)={\cal M}$, where ${\cal M}$ is a column vector whose first $N/2$ elements are zero and the next $N/2$ are $m$.  With careful use of the source term provided by the current-quark mass, one eliminates the trifurcation point, so that all three solution trajectories $H^{-1}({\cal M})$ become distinct but remain close.  A combination of iteration and the role change method can subsequently be used to find and track these solutions.

We note that in all cases when tracking a solution we ensure that Eq.\,\eqref{detneqzero} is satisfied at each point $X(\lambda)$ so we can be certain that no solution is missed.

\bibliographystyle{../../zProc/z10KITPC/h-physrev4}
\bibliography{../../CollectedBiB}
\end{document}